# High spatial resolution trace element determination of geological samples by laser ablation quadrupole plasma mass spectrometry: implications for glass analysis in volcanic products

Authors:

Maurizio Petrelli (maurizio.petrelli@unipg.it)[1*],

Kathrin Laeger (kathrin.laeger@studenti.unipg.it)[1],

Diego Perugini (diego.perugini@unipg.it)[1]

**Running Title:**

High spatial resolution trace element determination of geological samples by LA-ICP-MS

[1] Dept. of Physics and Geology, University of Perugia, Piazza dell'Università, 1, 06123 Perugia, Italy

[*] Corresponding author: phone +39 075 585 2607, fax +39 075 585 2603

**ABSTRACT**

Increasing the spatial resolution of Laser Ablation Inductively Coupled Plasma Mass Spectrometry (LA-ICP-MS) is a challenge in microanalysis of geological samples. Smaller sizes for the laser beam will allow for (1) high resolution determination of trace element compositions, (2) accurate estimation of crystal/melt partition coefficients, (3) detailed characterization of diffusion profiles, and (4) analysis of fine volcanic glasses. Here, we report about the figures of merit for LA-ICP Quadrupole MS down to a spatial resolution of 5 μm. This study highlights the possibility to achieve suitable limits of detection, accuracy and precision for geological samples even at spatial resolutions of the order of 5 μm. At a beam size of 15 μm precision (measured as one sigma) and accuracy (expressed as relative deviation from the reference value) are of the order of 10%. At a smaller beam size of 8um, precision decreases to 15% for concentration above 1.7 μg g$^{-1}$. As the beam size is decreased to ~5 μm, precision declines to about 15% and 20% for concentrations above 10 μg g$^{-1}$ using $^{42}$Ca and $^{29}$Si as internal standard, respectively. Accuracy is better or equal to 10% and 13 % at beam sizes of 15 and 10 μm respectively. When the spatial resolution is increased to 8 μm, accuracy remains better than 15% and 20% for $^{42}$Ca and $^{29}$Si as internal standard, respectively. We employed such high-resolution techniques to volcanic glasses in ash particles of the 2010 Eyjafjallajökull eruption. Our results are well consistent with the previously reported data obtained at lower spatial resolution, supporting the reliability of the method.

Keywords: LA-ICP-MS; High spatial resolution; Trace element determination; 5 micron; volcanic glasses

# 1. INTRODUCTION

Laser ablation inductively coupled plasma-mass spectrometry (LA-ICP-MS) is a powerful micro-analytical technique for trace element and isotopic ratio determinations. Since its introduction (Gray, 1985), it has been successfully applied in several fields like earth, natural and material sciences, biology, forensics, chemistry and industry (e.g. Norman et al., 1996, 1998; Jeffries et al., 1998; Becker and Dietze, 1999; Durrant, 1999; Günther et al., 1999; Horn et al., 2000; Sylvester, 2001; Košler et al., 2002; Tiepolo, 2003; Tiepolo et al., 2003; Petrelli et al., 2007, 2008; Alagna et al., 2008; Sylvester, 2008; Kil, 2011; Kil et al., 2011; Pozebon et al., 2014; Xiaoxia and Regelous, 2014; Kimura et al., 2015; Kil and Jung, 2015).

In a typical LA-ICP-MS trace element determination of geological material, a circular laser beam with a diameter ranging from 15 to 100 μm is focused on the area of interest, ablating this part of the sample. The ablated material is then transported to the ICP-MS where it is analysed. When working with laser beam diameters larger than 15 μm, lower limits of detection (LLD) are generally below than 1 μg g$^{-1}$ and 1 ng g$^{-1}$ for light-mass and heavy-mass elements, respectively (Durrant, 1999). Precisions and accuracies are generally equal to or better than 10% for all determined elements (Durrant, 1999).

By decreasing the laser beam diameter below 15 μm, less material reaches the ICP-MS, resulting in lower sensitivities, leading to a deterioration of limits of detection and larger analytical errors (Jeffries et al., 1995; Günther et al., 1996). Nevertheless, there is a large number of applications in earth sciences requiring laser beam diameters equal to or smaller than 15 μm: examples are crystal/melt partition coefficients (D$^{c/m}$) on experimental samples (e.g., Jenner et al., 1993; Petrelli et al., 2008), analyses of melt and fluid inclusions (e.g., Taylor et al., 1997; Günther et al., 1998, 2001; Halter et al., 2002) and analysis of glass shards from tephra deposits (e.g., Pearce et al., 2011).

Sector field ICP-MSs are generally the instruments of choice to perform successful LA-ICP-MS analyses with laser beam diameters less than or equal to 15 μm. This is due to their higher sensitivities relative to Quadrupole Mass spectrometers (QMS, Latkoczy and Günther, 2002; Paquette and Tiepolo, 2007; Pearce et al., 2011). It is important to note that recent quadrupole mass spectrometers have been considerably improved in terms of ion path geometry and plasma stability, resulting in a more efficient ion transmission through the instrument and better-quality neutrals removal (Bonta et al., 2015; Li et al., 2016; Petrelli et al., 2016). The result is an enhanced sensitivity and signal stability making these techniques suitable for high spatial resolution trace element determinations. It is notable that only a few studies report on the application of a LA-ICP-QMS with spatial resolution below 15 μm (e.g., Günther et al., 1996; Sigmarsson et al., 2011). The result is a lack of knowledge of the analytical figures of merit (*e.g.*, precision, limits of detection and accuracy) associated with such specific operating conditions.

Here we attempt to fill this gap reporting on the capabilities of a LA-ICP-QMS in trace element determinations of geological samples using laser beam diameters ranging from 15 to 5 μm. We address the figures of merit (LLD, precision and accuracy) of LA-ICP-QMS when operating at high spatial resolution mode and provide a case study of glasses in volcanic ash particles from the 2010 Eyjafjallajökull eruption (Iceland) using a spatial resolution of 8 μm. High spatial resolution LA-ICP-MS trace element analyses have many potential applications such as, for example: (1) high resolution determination of trace element compositions, (2) accurate estimation of crystal/melt partition coefficients, (3) detailed characterization of diffusion profiles, and (4) analysis of fine volcanic glasses.

## 2. ANALYTICAL METHODS

The used laser ablation device is a Teledyne / Photon Machine G2 equipped with a Two-Volume ANU (Australian National University) HelEx 2 cell.

The applied source is an ATL-I-LS-R solid-state triggered excimer 193 nm laser. It is characterized by a maximum stabilized energy output of 12 mJ with fluctuations, expressed as relative standard deviation (RSD), below 2%. The pulse duration was < 4 ns and the irradiance on the sample surface could be adjusted up to about 4 GW/cm$^2$. Frequencies could be varied from 1 Hz to 300 Hz.

The beam delivery apparatus was contained in a fully $N_2$-purged optical path to prevent formation of ozone and concomitant energy loss on the target. A rotating beam homogenizer improved the homogeneity of the laser beam through its surface; a fast-change mask varied the shape and dimension of the laser beam on the sample surface. The mask allowed, among others, nominal circular spots of 5, 8, 10 and 15 μm required for this study (Figure 1).

Figure 1 is about here

The ablation cell design is based on the extensively tested two volume HelEx cell developed at the Australian National University (Eggins et al., 1998; Woodhead et al., 2004) modified with an active flow in the funnel.

The ablation was performed under Helium (He) atmosphere to enhance the ablation performance, as well as to reduce particle deposition and inter-element fractionation (Eggins et al., 1998; Günther and Heinrich, 1999). Argon (Ar) and nitrogen (N) were added after the ablation cell in order to reduce perturbations on the plasma torch and to enhance sensitivity, respectively (Hu et al., 2008). Tubing length from the ablation cell to the ICP-MS was reduced at minimum (about 1 m) in order to reduce transport-related inter-element fractionation. A squid signal-smoothing device was positioned before the plasma torch in order to increase the stability of the signal (Müller et al., 2009).

The spectrometer is a quadrupole based Thermo Scientific iCAP-Q ICP-QMS. It is able to achieve sensitivities above $10^6$ cps/ng g$^{-1}$ for $^{110}$In and $^{238}$U when operating in spray chamber liquid introduction mode while maintaining relative standard deviations below 2% measured on 10 separate acquisitions of one minute each. Under these operating conditions, oxide formations, measured as CeO$^+$/Ce$^+$ and double charged ions evaluated as Ba$^{++}$/Ba$^+$, are below 2% and 3%, respectively.

LA-ICP-MS operating conditions were optimized before each analytical session on a continuous ablation of NIST SRM 612 reference material glass in order to provide the maximum signal intensities and stabilities for the ions of interest while suppressing oxides formation. The latter were monitored keeping the ThO$^+$/Th$^+$ ratio below 0.5%. The U/Th ratio was also monitored and maintained close to 1.

The stability of the system was evaluated daily on $^{139}$La, $^{208}$Pb, $^{232}$Th and $^{238}$U by a short-term stability test. It consisted of 5 acquisitions (one minute each) on a linear scan of NIST SRM 612 reference material glass.

The analytical protocol for high spatial resolution trace element determination consists in the analysis of 10-15 unknown samples bracket by four acquisitions of the NIST SRM 610 reference material. The diameter of the laser beam was varied between 15 and 5 µm for the unknown samples whereas the reference material was always analyzed at a beam size of 15 µm in order to improve the counting statistics on calibration analyses (Petrelli et al., 2007). The USGS BCR2G (Wilson, 1997) was used here for the estimation of the figures of merit and as Quality Control in routine analysis. Data reduction was performed using the protocol reported in Longerich et al. (1996). $^{42}$Ca and $^{29}$Si were used as internal standards. A summary of the operating conditions of the LA-ICP-MS is reported in Table 1.

Table 1 is about here

Major elements on unknown samples were analyzed with an Electron Probe Micro-Analyzer (hereafter EPMA). EPMA analyses were performed with a Jeol-JXA8200 WDS using an accelerating voltage of 15 kV and an electric current of 10 nA. A defocused electronic beam of 5 μm was used with a counting time of 5s on background and 15s on peak. The following standards were used for the different chemical elements: jadeite (Si and Na), corundum (Al), forsterite (Mg), andradite (Fe), rutile (Ti), orthoclase (K), barite (Ba), celestine (S), fluorite (F), apatite (P and Cl) and spessartina (Mn). Sodium and potassium were analyzed first in order to minimize possible volatilization effects.

## 3. RESULTS

### 3.1. Lower Limits of Detection (LLD) for the Internal Standard

In LA-ICP-MS, LLDs depend on both the background and sensitivity of the instrumentation. In detail, they are a function of the amount of material reaching the ICP-MS and therefore, they are function, among the other parameters, of the laser beam diameter (Günther et al., 1999). Given that the data reduction protocol adopted here (Longerich et al., 1996) requires the use of an internal standard, the first step in the assessment of the instrumentation for high spatial resolution trace element determinations consists in the evaluation of LLD for $^{29}$Si and $^{42}$Ca. LLDs for $^{29}$Si and $^{42}$Ca are estimated according to Perkins and Pearce (1995) at beam sizes of 15, 10, 8 and 5 μm and the results are reported in Figure 2. Lower limits of quantification (LLQ) evaluated as LLQ = 3.333 x LLD (Perkins and Pearce, 1995) are also reported in Figure 2.

LLQs progressively increase from 4.5 to 22.8 wt.% for $SiO_2$ and from 1.1 to 6.2 wt.% for CaO as the laser beam diameter is decreased from 15 to 5 μm.

Figure 2 is about here

### 3.2. LLDs for Unknown Elements

LLDs calculated following Longerich et al. (1996) for selected elements at different spot sizes are reported in Table 2 and displayed in Figure 3. LLDs vary between 0.006 and 2 μg g$^{-1}$, 0.02 and 3 μg g$^{-1}$, 0.03 and 4 μg g$^{-1}$ and 0.07 and 9 μg g$^{-1}$ at laser beam diameters of 15, 10, 8, and 5 μm relative to the USGS BCR2G reference material at 8 Hz and 4.5 J/cm$^2$.

Table 2 is about here

Reported LLDs are of the same magnitude of those reported by Tiepolo et al. (2003) and Pearce et al. (2011) using a sector field ICP-MS coupled with a solid state 213 nm and an Excimer 193 nm laser ablation system, respectively.

Figure 3 is about here

### 3.3. Precision and Accuracy

Table 3 and Table 4 report the values of precision and accuracy evaluated for spot sizes of 15, 10, 8 and 5 μm and acquired in two different analytical sessions using $^{42}$Ca and $^{29}$Si as internal standard, respectively. Precision is evaluated as Relative Standard Deviation (RSD %, 1 sigma) whereas accuracy is evaluated as Relative Deviation from the Reference Value (RDRV %).

Table 3 is about here

Figure 4 shows the Relative Standard Deviation in % (RSD %, n=6) plotted against the concentration of the analysed isotopes. At a beam size of 15 µm (Figure 4a), precision is equal to or better than 10% for all the elements. Above 10 µg g$^{-1}$, precision improves to 6% and, above 100 µg g$^{-1}$ it is better than 5%. At a beam size of 10 µm (Figure 4b), precision is equal to or better than 10% for concentrations above 1.7 µg g$^{-1}$. For concentrations below 1.7 µg g$^{-1}$ precision values progressively degrade to more than 20%. At a beam size of 8 µm (Figure 4c), precision is better than 15% for concentrations above 1.7 µg g$^{-1}$. For concentrations below 1.7 µg g$^{-1}$ precision values progressively deteriorate to ca. 40%. There are no significant differences in the use of $^{29}$Si and $^{42}$Ca as internal standard in terms of precision at beam sizes of 15, 10 and 8 µm. At a beam size of 5 µm (Figure 4d), precision is better than 15% and 17% for concentrations above 10 µg g$^{-1}$ using $^{29}$Si and $^{42}$Ca as internal standard, respectively. Decreasing the concentration of the analysed isotope below 10 µg g$^{-1}$, precision values progressively deteriorate to 50-60% using either $^{29}$Si or $^{42}$Ca as internal standard.

Accuracy values are reported on Table 3, Table 4 and Figure 5. At a beam size of 15 µm, accuracy values are equal to or better than 10% using either $^{29}$Si or $^{42}$Ca as internal standard (Figure 5). At a beam size of 10 µm, accuracy is still better than or equal to 10% for $^{42}$Ca as internal standard; accuracy is better then 13% when $^{29}$Si is used as internal standard (Figure 5). At a beam size of 8 µm, accuracies are better than 15% and 20% using $^{42}$Ca and $^{29}$Si as internal standard, respectively (Figure 5). At a beam size of 5 µm, accuracies are of the order of 25-30% using either $^{29}$Si or $^{42}$Ca as internal standard (Table 3, 4). In order to evaluate the relationship between the accuracy and the concentration at a beam size of 5 µm, absolute values of RDRV are plotted against concentration of the analyzed isotopes in Figure 6. The figure shows that accuracies are ca. 30% for concentration below ca. 2 and 5 µg g$^{-1}$ for $^{42}$Ca and $^{29}$Si as internal standard, respectively. Increasing the concentration, the accuracy improves to 15% and 20% with $^{42}$Ca and $^{29}$Si as internal standard, respectively.

Comparing the use of $^{42}$Ca and $^{29}$Si as internal standard in high spatial resolution analysis, no evidence of inter-element fractionation emerges operating with a beam diameter of 15 µm. Reducing the laser beam diameter to 10 and 8 µm, slight under-estimation of the real values occurs when Si is used as internal standard (Figure 5b).

Figure 6 is about here

### 3.4. Spot Size and Elemental fractionation

Fryer et al. (1995) defined the elemental fractionation as non-stoichiometric ablation that involves non-sample related changes in element intensities with time. Elemental fractionation is one of the major sources of inaccuracy in LA-ICP-MS analysis and is a function of several parameters such as laser wavelength, fluence, focus position, focal length of the objective, repetition rate, crater size, depth of the ablation and sample material (Günther 2001). Generally, in trace element determinations, elemental fractionation is minimized by fixing the laser operating conditions for reference samples and unknowns and using an internal standard with an elemental fractionation behavior similar to the elements of interest.

We studied the elemental fractionation effects related to the use of different spot sizes for calibrators and unknowns evaluating the accuracies reported for the USGS BCR2G. The occurrence of fractionation effects associated with the use of different spot sizes during the analysis will result in a deviation of the accuracy values for the analyses at beam sizes of 10, 8 and 5 μm when compared with the results obtained using a 15 micron spot size.

Figure 7 is about here

Figure 7 illustrates box plot distributions of the relative deviation of the accuracies at different spot sizes (5, 8 and 10 μm) considering the median value of accuracy obtained at a beam size of 15 microns as a reference value. In the figure the red lines correspond to the median of the distribution and the rectangles represent the middle half of the data; they range from the 25th percentile to the 75th percentile. The vertical dashed lines represent the range from the minimum to the maximum values excluding the outliers. In Figure 7, a perfectly unbiased distribution would show a Gaussian distribution with a median equal to zero.

Figure 7 highlights that the accuracy distributions at 10 and 8 microns are statistically unbiased compared to the results at a beam size of 15 microns for both Ca and Si as internal standard. Results at a beam size of 5 microns shows a statistically significant bias when Si is utilized as the internal standard (Figure 7b; lower values). The accuracies obtained using Ca as internal standard, although widespread (i.e. low accuracies) are not statistically biased from the values obtained at a beam size of 15 microns (Figure 7a).

### 3.5. Example Application

The April–May 2010 eruption of the Eyjafjallajökull volcano (EFJ, Iceland) represents an example of explosive activity dominated by a nearly continuous injection of ash into the atmosphere. Studied ash particles are derived from the last days of the eruption. Ash samples are constituted by phenocrysts hosted in a heterogeneous microcrystalline to glassy groundmass. A preliminary characterization by EPMA indicates a trachyandesite to trachydacite composition of the glassy portion with $SiO_2$ and $CaO$ contents ranging from 60.47 to 69.54 wt.% and 1.26 to 3.97 wt.%, respectively (Table 5).

Figure 8 is about here

In order to test the analytical protocol described here, the glass portion of the ash was analysed by using a spot size of 8 μm (Figure 8). $^{29}Si$ was used as internal standard as the concentration of $^{42}Ca$ at a beam size of 8 μm was below the LLQ. Results are reported in Table 5 and plotted in Figures 9 and 10. Further, as a reference, the trace element composition of Sigmarsson et al. (2011) and Borisova et al. (2012) glass portions of tephra erupted in an earlier stage between 14.04. - 20.05.10 are also reported.

Table 5 is about here

Figure 9 is about here

Figure 10 is about here

Bivariate diagrams of trace elements plotted using Zr as differentiation index (Figure 9) show that the obtained results agree well with literature data and support the applicability of the data in real case studies. In detail, for all the diagrams in Figure 9, the results of our analysis define the same trends reported in literature (Sigmarsson et al., 2011; Borisova et al., 2012). The Rare Earth Element (REE) spider diagram (Figure 10) normalized to chondrite values provided by Sun and McDonough (1989) also shows a good matching with data reported by Sigmarsson et al. (2011) and Borisova et al. (2012).

## 4. DISCUSSION AND CONCLUSIONS

Increasing the spatial resolution of LA-ICP-MS is a challenge in micro-analytical trace element determination of geological samples because of the continuous request of progressively more detailed studies at increasing magnifications. Here, we demonstrated the possibility to achieve suitable limits of detection for geological samples even at spatial resolutions of the order of 5 μm by using quadrupole mass spectrometers. In this regard, the choice of the most suitable internal standard is of crucial importance. In fact, it should be noted that LLQ for $SiO_2$ is always lower than the typical content of igneous systems where $SiO_2$ generally ranges from 45 wt.% to 75 wt.%. In contrast, CaO should not be used with resolution below 10 μm for high-silica igneous rock where CaO is generally lower than 3.0 wt.%.

Results also show that it is possible to achieve precisions and accuracies of the order of 10% at a beam size of 10 μm. At higher resolution precisions decrease progressively but they are still better than 15% at a beam size of 8 μm for concentrations above 1.7 $\mu g\,g^{-1}$. At a beam size of 5 μm, precision is better than 15% and 17% for concentrations above 10 $\mu g\,g^{-1}$ using $^{29}$Si and $^{42}$Ca as internal standard, respectively. Accuracies are better or equal to 10% and 13% at beam sizes of 15 and 10 μm, respectively. Increasing the spatial resolution to a beam size of 8 μm, accuracy is still better than 15% and 20% using $^{42}$Ca and $^{29}$Si as internal standard, respectively. At a beam size of 5 μm, the accuracies are of the order of 20-30%.

Comparing results on the use of $^{29}$Si and $^{42}$Ca as internal standard in term of accuracy, it emerges that increasing the resolution to a beam size of 5 μm causes a negative bias for $^{29}$Si (Figure 5b). This analytical bias has been also observed by Pearce et al. (2011) and it could be the result of non-stoichiometric effects occurring during the ablation process, transport and ionization of the sample in the plasma, collectively referred to as "fractionation" (Pearce et al., 2011). These effects have been already reported and discussed widely in literature (Guillong and Günther, 2002; Günther and Koch, 2008; Gaboardi and Humayun, 2009). A reduced fractionation is observed when using $^{42}$Ca as internal standard at beam sizes of 10 and 8 µm.

It is significant to note that obtained precisions and accuracies are suitable for most geological application down to a beam size of 8 µm.

The application of 8 µm spatial resolution to volcanic glasses provided results that agree well with literature data obtained at lower resolutions of 40 and 11 µm (Sigmarsson et al., 2011; Borisova et al., 2012).

Accuracy and precision at a beam size of 5 µm tend to decrease but they are still suitable for several applications. A possible example is the study of ultrahigh-pressure (UHP) multiphase inclusions (Ferrando et al., 2009). Up to now, these inclusions have been studied by quantifying the bulk concentration of the about 20 µm assemblage. By increasing the spatial resolution to a beam size of 5 µm it could be possible, in principle, to analyse each single phase of about 5 µm length. Another potential application is the trace element imaging of melt inclusions. Very high spatial resolution (5 µm) might also be a useful tool to evaluate the interaction between the inclusion and the host material.

As a concluding remark, we wish to highlight that the use of quadrupole mass spectrometers, as demonstrated in the present study, is a valid tool in high resolution (down to 8 microns) trace element determinations opening new perspectives to many quadrupole-based LA-ICP-MS laboratories.

## ACKNOWLEDGEMENTS

This study was supported by the CHRONOS project funded under the framework "FP7-IDEAS-ERC" (ERC-CG-2013-PE10, 612776), by the VERTIGO Initial Training Network funded under "FP7-PEOPLE-2013-ITN" (607905) and by the PRIN project 2010-11. We acknowledge the Editor (K. Min) and two anonymous reviewers for the constructive comments and suggestion. We also acknowledge Rebecca Astbury for the proofreading of the final version of the manuscript.

File Rep. 98 p.

**Tables and Figures**

Table 1

ICP-MS and Laser Ablation operating conditions.

Table 2

Lower limits of detection (LLD in $\mu g\,g^{-1}$) at different spatial resolutions calculated following Longerich et al., (1996).

Table 3

Precision (expressed as Relative Standard Deviation, RSD) and accuracy (expressed as Relative deviation from the reference value, RDRV) calculated at different spatial resolution on the USGS BCR2G reference material (Wilson, 1997) and using $^{42}$Ca as internal standard.

Table 4

Precision (expressed as Relative Standard Deviation, RSD) and accuracy (expressed as Relative Deviation from the Reference Value, RDRV) estimated at different spatial resolution on the USGS BCR2G reference material (Wilson, 1997) and using $^{29}$Si as internal standard.

Table 5

Major and trace element concentrations in $\mu g\,g^{-1}$ of trachyandesite to trachedacite glasses in ash particles of 2010 Eyafjallajökull eruption at spatial resolution of 8 $\mu m$.

Figure 1

Ablation craters on the NIST SRM610 reference material at beam sizes of 10, 8 and 5 µm. Image acquired by optical microscope.

Figure 2

Lower limits of detection (LLD, white symbols) and lower limits of quantification (LLQ, black symbols) for $^{29}$Si (squares) and $^{42}$Ca (circles) plotted against the laser beam diameter (in µm).

Figure 3

Lower limits of detection (LLD) at beam sizes of 15, 10, 8 and 5 µm for all relevant elements analyzed on the USGS BCR2G (Wilson, 1997) reference material.

Figure 4

Precisions (RSD in %) reported against the concentrations of the analysed isotope at different spatial resolutions (15, 10, 8 and 5 µm) calculated on the USGS BCR2G reference material (Wilson, 1997): (a) 15 µm, (b) 10 µm, (c) 8 µm and (d) 5 µm using $^{42}$Ca (circles) and $^{29}$Si (squares) as internal standard (IS).

Figure 5

Accuracies (RDRV in %) at different spatial resolutions (15, 10, and 8 µm) obtained using $^{42}$Ca (a) and $^{29}$Si (b) as internal standard (IS).

Figure 6

Accuracies in RDRV in % reported as the concentration of the analysed isotopes for both [42]Ca (black circles) and [29]Si (white squares) as internal standard (IS). Dashed and continuous lines represent upper values of RDRV for [42]Ca and [29]Si, respectively.

Figure 7

Box plots reporting the deviations of the accuracies form the median value obtained at 15 µm: (a) [42]Ca as internal standard; (b) [29]Si as internal standard.

Figure 8

Back Scattering Electrons (BSE) image of two Eyjafjallajökull ash grains evidencing three 8 µm laser ablation craters. Black circles with an outer diameter of 8 µm are overlain to the laser ablation craters to highlight the spatial resolution of the analysis.

Figure 9

Selected trace elements (La, Rb, Ta and Th) plotted versus Zr for Eyafjallajökull glasses (white circles). Data from Borisova et al. (2012) (grey pentagons) and Sigmarsson et al. (2011) (grey crosses) are also reported.

Figure 10

Condrite-normalized rare earth element (REE) spider diagram (Sun and McDonough, 1989) for Eyjafjallajökull glasses. Data from Borisova et al. (2012) and Sigmarsson et al. (2011) are also reported as a grey area.

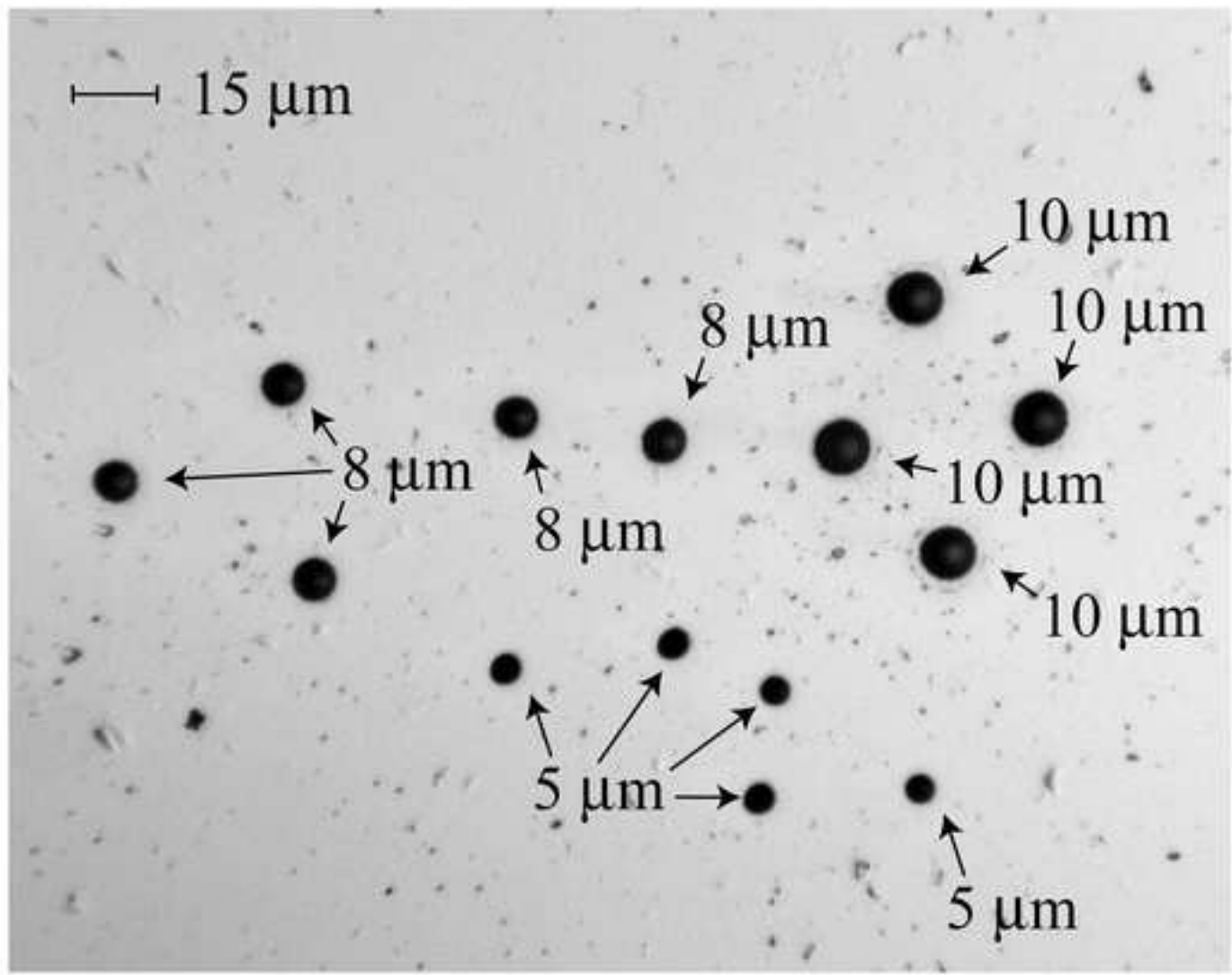

Figure 1

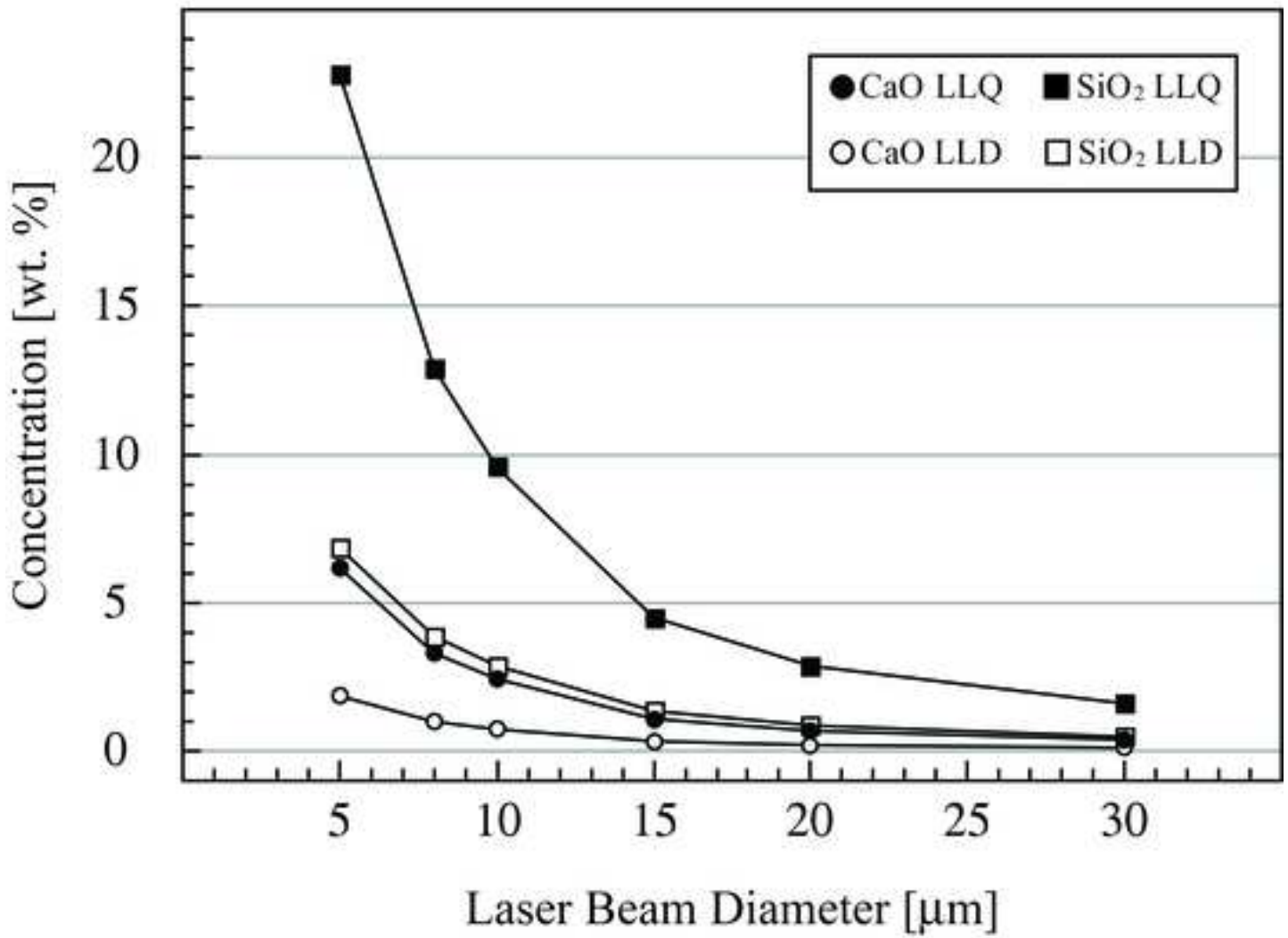

Figure 2

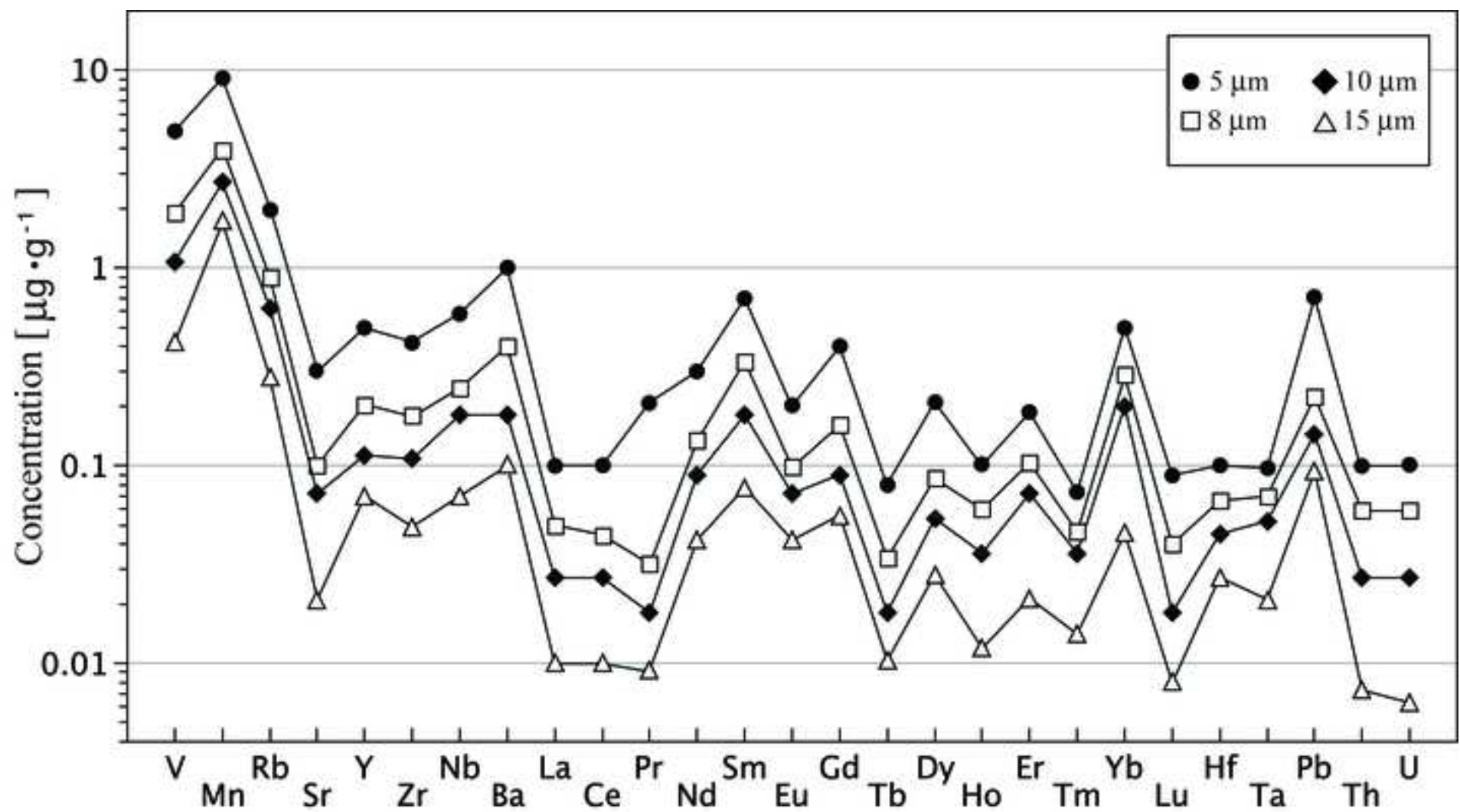

Figure 3

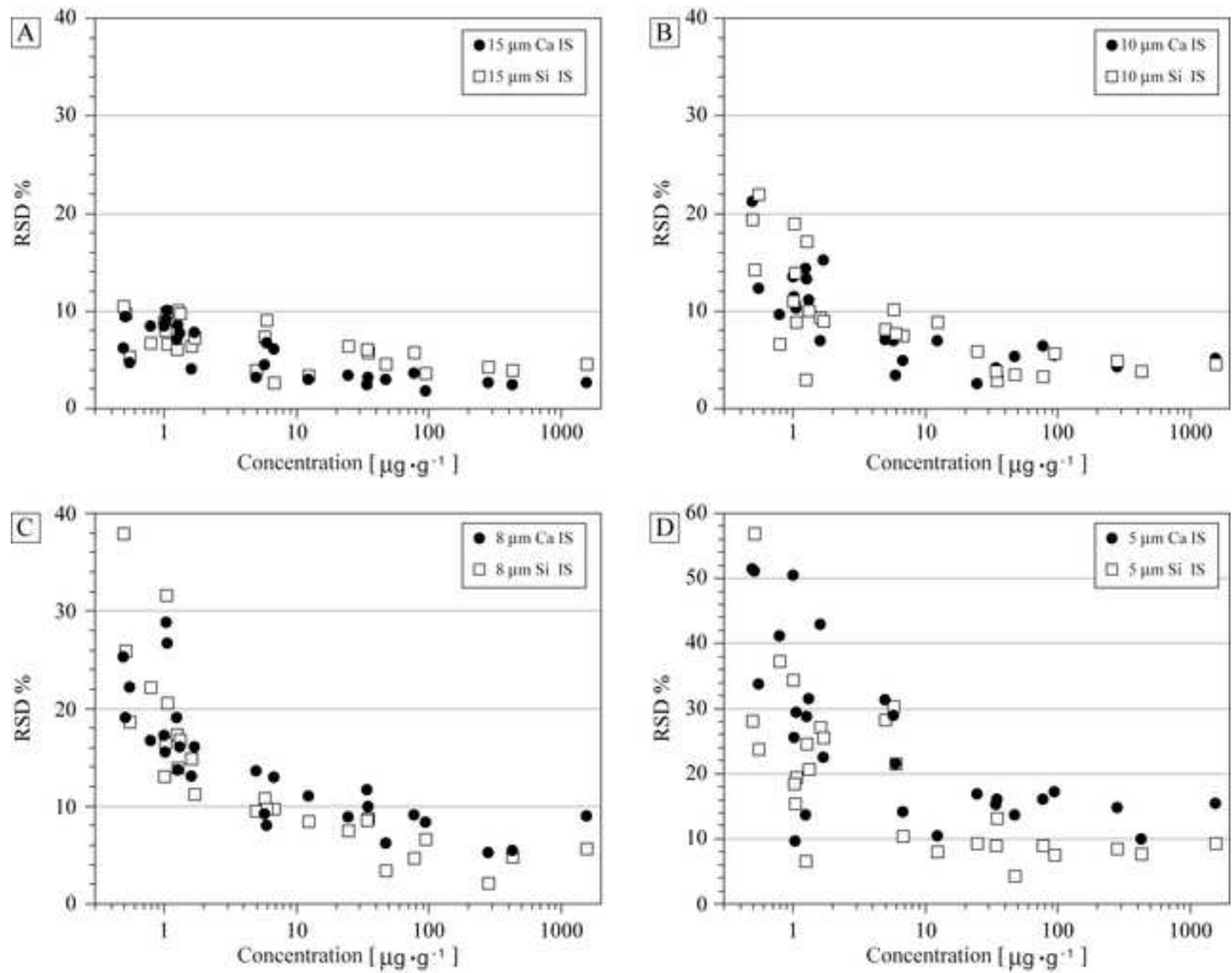

Figure 4

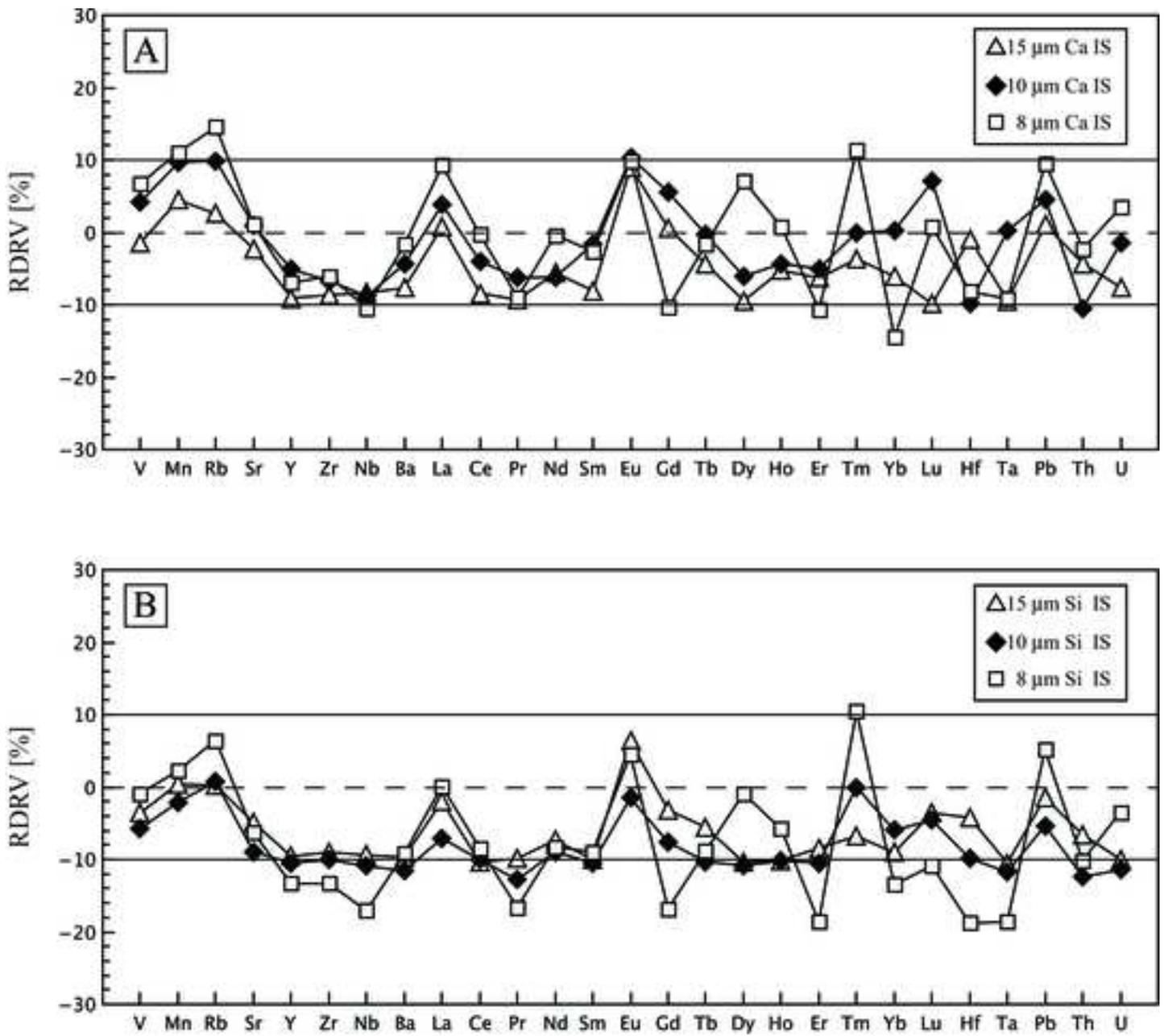

Figure 5

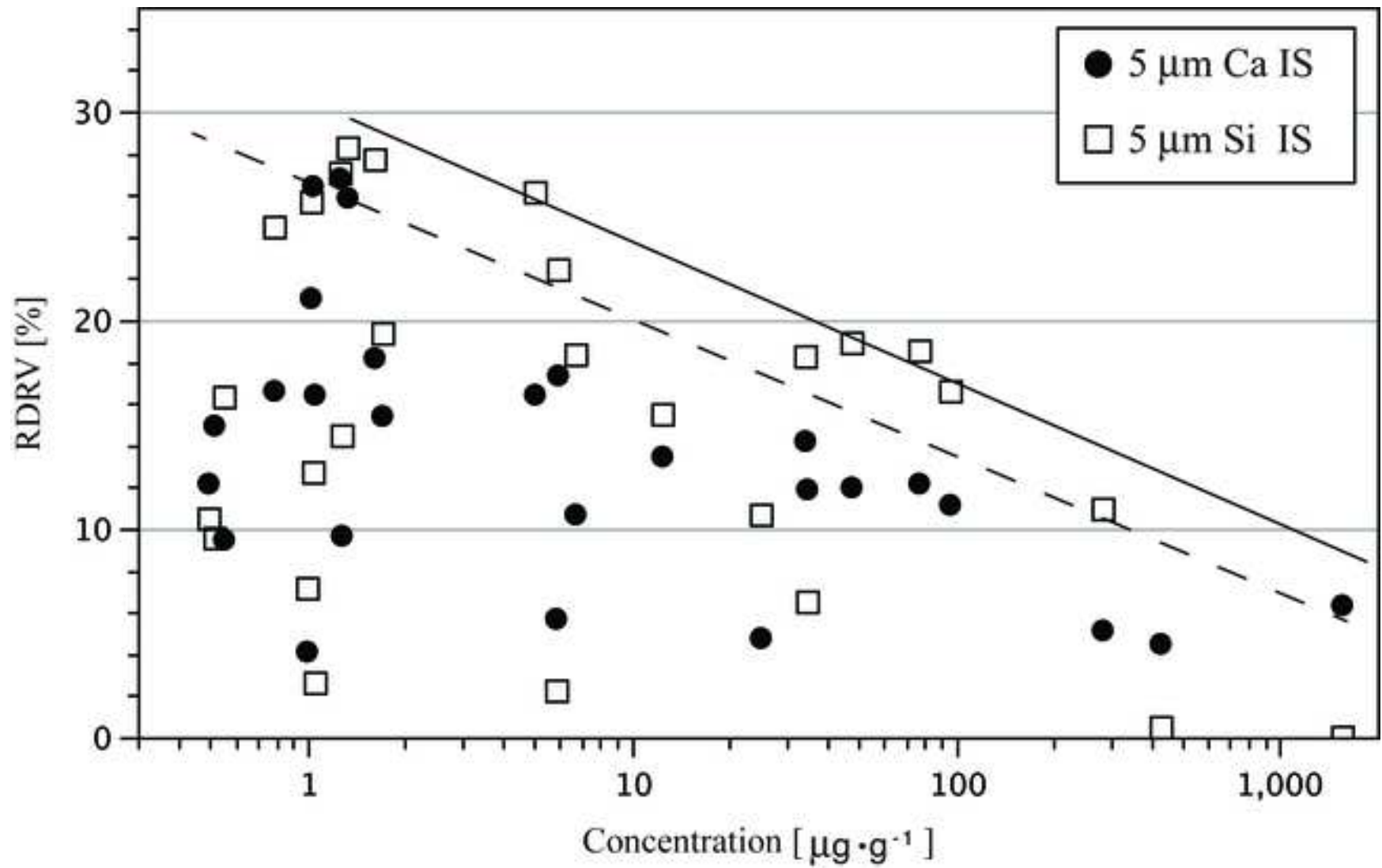

Figure 6

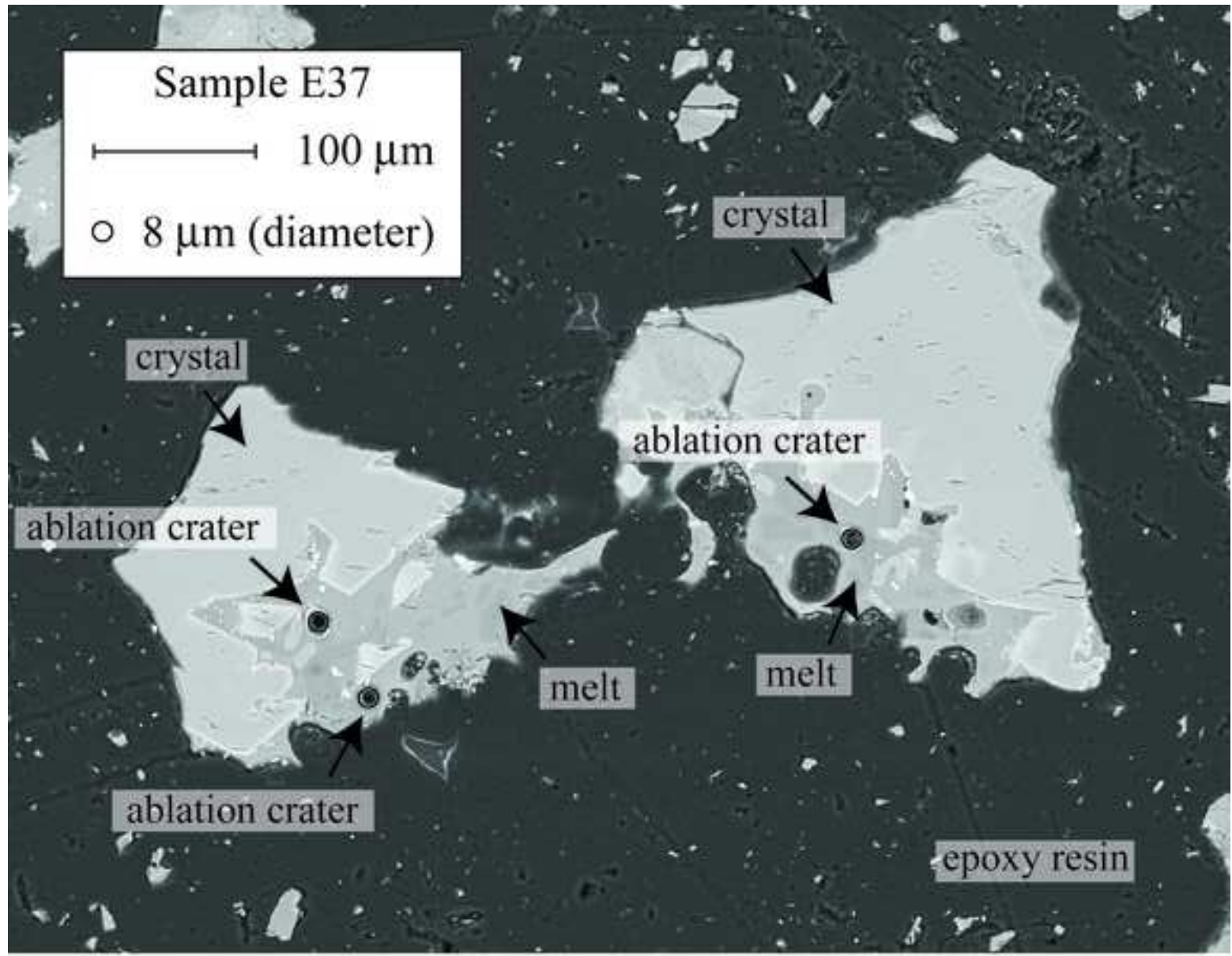

Figure 7

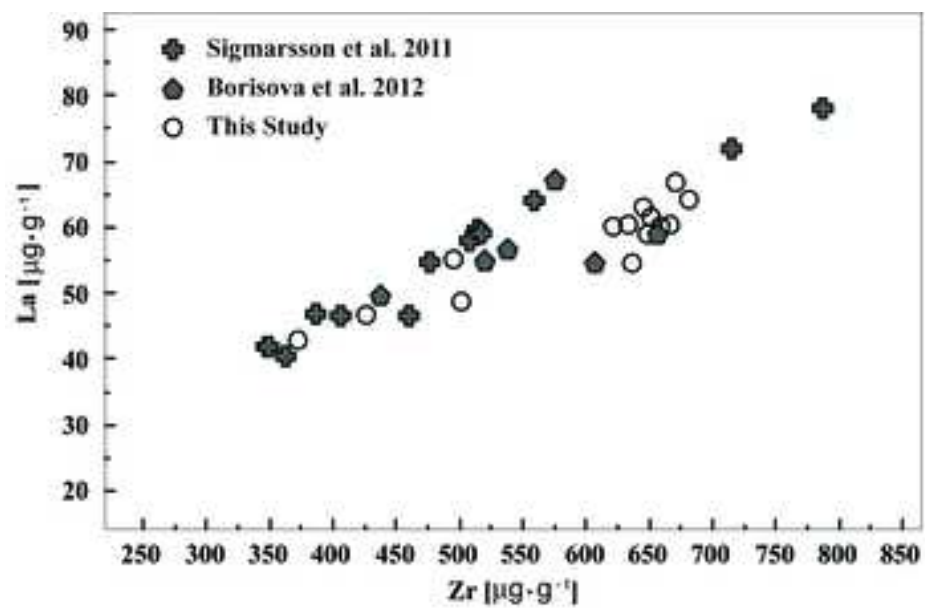
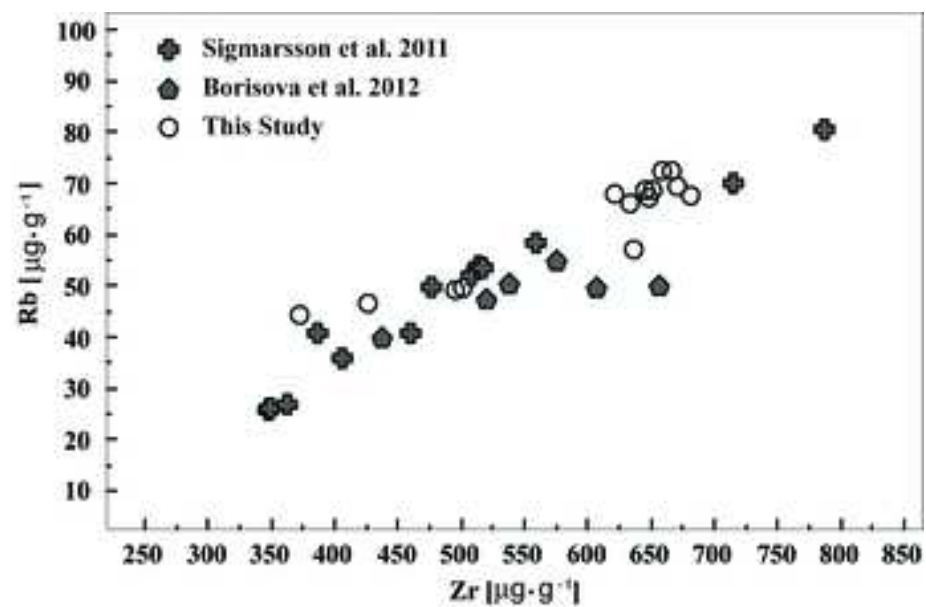
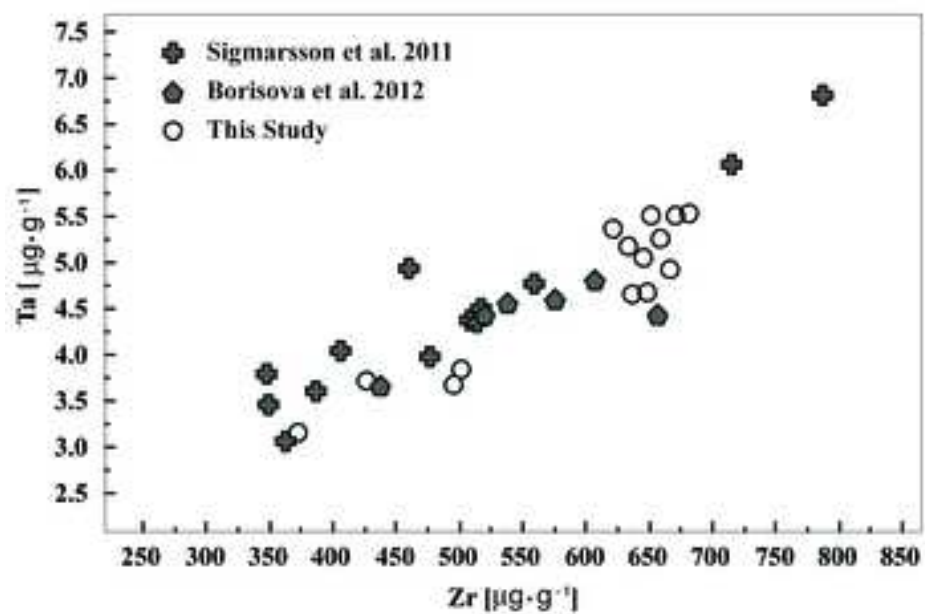
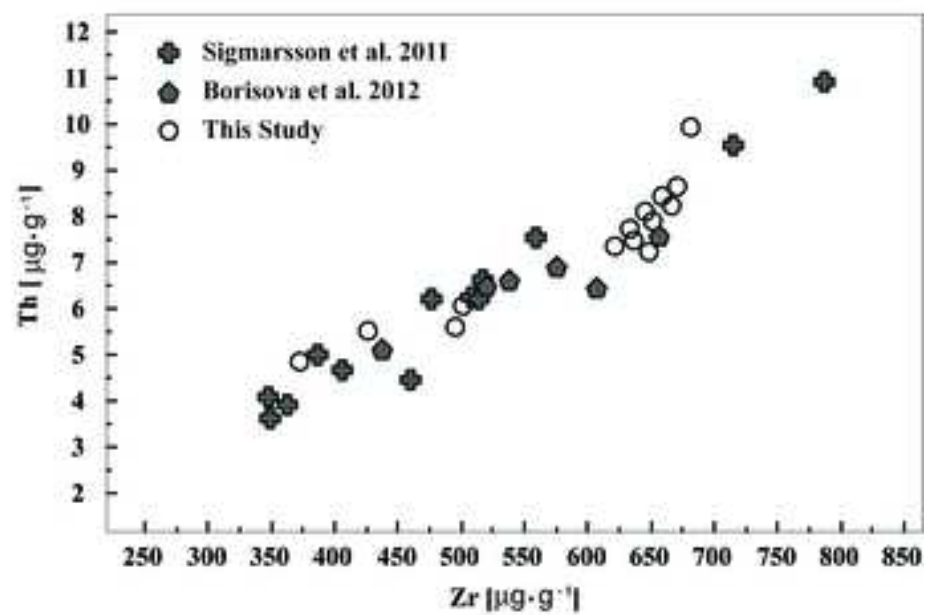

Figure 8

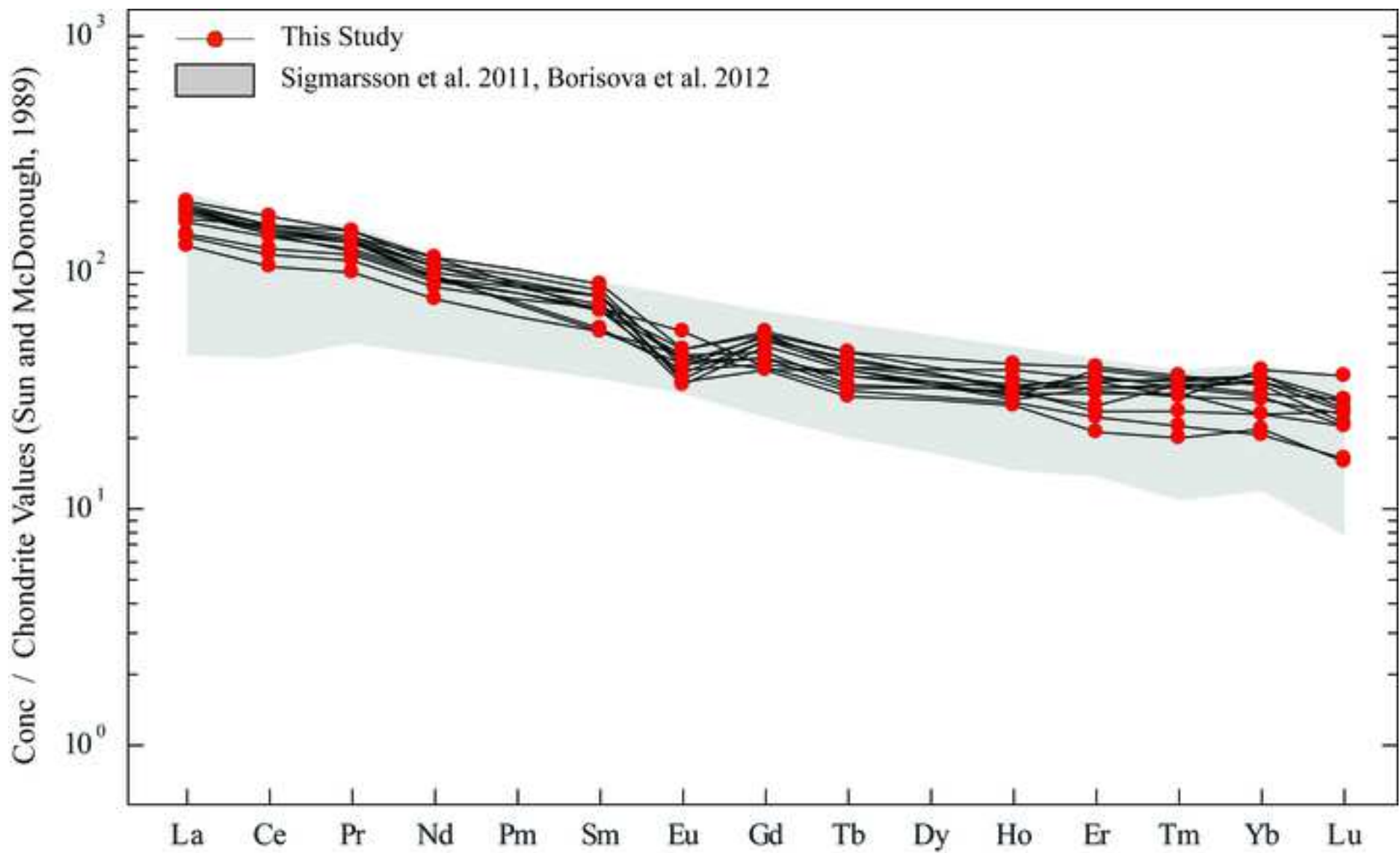

Figure 9

| ICP-MS: | Thermo Fisher Scientific iCAP Q |
|---|---|
| RF power | 1400 - 1550 W |
| Sampling depth | 4.5 - 5.5 mm |
| Carrier gas flow (Ar) | 0.55 - 0.75 l/min |
| Coolant gas flow  (Ar) | 14 l/min |
| Auxiliary gas flow  (Ar) | 0.8 l/min |
| Dwell time/mass | 15 ms |
| Masses (m/z) | $^{29}$Si, $^{42}$Ca, $^{51}$V, $^{55}$Mn, $^{85}$Rb, $^{88}$Sr, $^{89}$Y, $^{90}$Zr, $^{93}$Nb. $^{137}$Ba, $^{139}$La, $^{140}$Ce, $^{141}$Pr, $^{146}$Nd, $^{147}$Sm, $^{153}$Eu, $^{157}$Gd, $^{159}$Tb, $^{163}$Dy, $^{165}$Ho, $^{166}$Er, $^{169}$Tm, $^{173}$Yb, $^{175}$Lu, $^{178}$Hf, $^{181}$Ta, $^{208}$Pb, $^{232}$Th, $^{238}$U |
| Sampler,  skimmer | Ni (with high sensitivity insert) |

| LA system: | Teledyne/Photon Machine G2 |
|---|---|
| Fluence on target | 0 - 12 J/cm$^2$ |
| ThO+/Th+ | < 0.5 % |
| He gas flow - cell | 0.4-0.8 l/min |
| He gas flow - funnel | 0.1-0.4 l/min |
| N2 gas flow | 4-5 ml/min |
| Laser repetition rate | 10 Hz |
| Transport tubing | < 1.0 m |
| Spot Sizes | 5, 8, 10, 15   m |

Table 1

| Element | Diameter of the spot size | | | |
|---|---|---|---|---|
| | 5 micron | 8 | 10 | 15 |
| V | 5 | 2 | 1 | 0.4 |
| Mn | 9 | 4 | 3 | 2 |
| Rb | 2 | 0.9 | 0.6 | 0.3 |
| Sr | 0.3 | 0.1 | 0.07 | 0.02 |
| Y | 0.5 | 0.2 | 0.1 | 0.07 |
| Zr | 0.4 | 0.2 | 0.1 | 0.05 |
| Nb | 0.6 | 0.2 | 0.2 | 0.07 |
| Ba | 1 | 0.4 | 0.2 | 0.1 |
| La | 0.1 | 0.05 | 0.03 | 0.01 |
| Ce | 0.1 | 0.04 | 0.03 | 0.01 |
| Pr | 0.2 | 0.03 | 0.02 | 0.009 |
| Nd | 0.3 | 0.1 | 0.09 | 0.04 |
| Sm | 0.7 | 0.3 | 0.2 | 0.08 |
| Eu | 0.2 | 0.1 | 0.07 | 0.04 |
| Gd | 0.4 | 0.2 | 0.09 | 0.06 |
| Tb | 0.08 | 0.03 | 0.02 | 0.01 |
| Dy | 0.2 | 0.09 | 0.05 | 0.03 |
| Ho | 0.1 | 0.06 | 0.04 | 0.01 |
| Er | 0.2 | 0.1 | 0.07 | 0.02 |
| Tm | 0.07 | 0.05 | 0.04 | 0.01 |
| Yb | 0.5 | 0.3 | 0.1 | 0.05 |
| Lu | 0.09 | 0.04 | 0.02 | 0.008 |
| Hf | 0.1 | 0.07 | 0.05 | 0.03 |
| Ta | 0.1 | 0.07 | 0.05 | 0.02 |
| Pb | 0.7 | 0.2 | 0.1 | 0.09 |
| Th | 0.1 | 0.06 | 0.03 | 0.007 |
| U | 0.1 | 0.06 | 0.03 | 0.006 |

Table 2

Table 3

| Element | 15 micron | | | | 10 micron | | | | 8 micron | | | | 5 micron | | | | REF |
|---|---|---|---|---|---|---|---|---|---|---|---|---|---|---|---|---|---|
| | mean (n=6) | 1 | RSD % | RDRV % | mean (n=6) | 1 | RSD % | RDRV % | mean (n=6) | 1 | RSD % | RDRV % | mean (n=6) | 1 | RSD % | RDRV % | |
| V | 419 | 10 | 2 | -1 | 443 | 17 | 4 | 4 | 454 | 25 | 5 | 7 | 444 | 44 | 10 | 5 | 425 |
| Mn | 1619 | 43 | 3 | 4 | 1700 | 87 | 5 | 10 | 1722 | 156 | 9 | 11 | 1648 | 254 | 15 | 6 | 1550 |
| Rb | 49 | 2 | 3 | 3 | 53 | 2 | 4 | 10 | 55 | 6 | 10 | 15 | 54 | 9 | 16 | 12 | 48 |
| Sr | 334 | 9 | 3 | -2 | 345 | 15 | 4 | 1 | 346 | 18 | 5 | 1 | 324 | 48 | 15 | -5 | 342 |
| Y | 30.9 | 0.7 | 2 | -9 | 32 | 1 | 4 | -5 | 32 | 4 | 12 | -7 | 29 | 4 | 15 | -14 | 34 |
| Zr | 168 | 3 | 2 | -9 | 172 | 9 | 5 | -7 | 173 | 14 | 8 | -6 | 163 | 28 | 17 | -11 | 184 |
| Nb | 11.3 | 0.3 | 3 | -8 | 11.2 | 0.8 | 7 | -9 | 11 | 1 | 11 | -10 | 11 | 1 | 11 | -14 | 12.3 |
| Ba | 631 | 23 | 4 | -8 | 653 | 42 | 6 | -4 | 672 | 61 | 9 | -2 | 600 | 96 | 16 | -12 | 683 |
| La | 24.9 | 0.8 | 3 | 1 | 25.7 | 0.7 | 3 | 4 | 27 | 2 | 9 | 9 | 24 | 4 | 17 | -5 | 24.7 |
| Ce | 49 | 1 | 3 | -8 | 51.2 | 2.7 | 5 | -4 | 53 | 3 | 6 | 0 | 47 | 6 | 14 | -12 | 53.3 |
| Pr | 6.1 | 0.4 | 6 | -9 | 6.3 | 0.3 | 5 | -6 | 6.1 | 0.8 | 13 | -9 | 6.0 | 0.8 | 14 | -11 | 6.7 |
| Nd | 27.3 | 0.9 | 3 | -6 | 27 | 2 | 7 | -6 | 29 | 4 | 14 | 0 | 24 | 8 | 31 | -16 | 28.9 |
| Sm | 6.1 | 0.5 | 8 | -8 | 6.5 | 0.9 | 13 | -2 | 6 | 1 | 17 | -3 | 7 | 3 | 51 | 4 | 6.59 |
| Eu | 2.1 | 0.2 | 9 | 9 | 2.2 | 0.2 | 11 | 10 | 2.2 | 0.6 | 29 | 10 | 2.5 | 0.2 | 10 | 26 | 1.97 |
| Gd | 6.8 | 0.7 | 10 | 1 | 7.1 | 0.7 | 10 | 6 | 6 | 2 | 27 | -10 | 8 | 2 | 29 | 16 | 6.71 |
| Tb | 0.98 | 0.09 | 9 | -4 | 1.0 | 0.1 | 11 | 0 | 1.0 | 0.2 | 16 | -2 | 0.8 | 0.2 | 26 | -21 | 1.02 |
| Dy | 5.8 | 0.2 | 4 | -9 | 6.1 | 0.4 | 7 | -4 | 6.9 | 0.9 | 13 | 7 | 5 | 2 | 43 | -18 | 6.44 |
| Ho | 1.2 | 0.1 | 9 | -5 | 1.2 | 0.2 | 13 | -4 | 1.3 | 0.2 | 14 | 1 | 1.1 | 0.3 | 29 | -10 | 1.27 |
| Er | 3.5 | 0.2 | 7 | -6 | 3.5 | 0.5 | 14 | -5 | 3.3 | 0.6 | 19 | -11 | 2.7 | 0.4 | 14 | -27 | 3.7 |
| Tm | 0.49 | 0.05 | 9 | -4 | 0.51 | 0.07 | 14 | 0 | 0.6 | 0.1 | 19 | 11 | 0.4 | 0.2 | 51 | -15 | 0.51 |
| Yb | 3.2 | 0.2 | 5 | -6 | 3.4 | 0.4 | 12 | 0 | 2.9 | 0.6 | 22 | -14 | 3 | 1 | 34 | -10 | 3.39 |
| Lu | 0.45 | 0.03 | 6 | -10 | 0.5 | 0.1 | 21 | 7 | 0.5 | 0.1 | 25 | 1 | 0.4 | 0.2 | 51 | -12 | 0.503 |
| Hf | 4.8 | 0.4 | 8 | -1 | 4.4 | 0.5 | 11 | -10 | 4.4 | 0.7 | 16 | -8 | 4 | 1 | 32 | -26 | 4.84 |
| Ta | 0.71 | 0.06 | 9 | -9 | 0.78 | 0.08 | 10 | 0 | 0.7 | 0.1 | 17 | -9 | 0.7 | 0.3 | 41 | -17 | 0.78 |
| Pb | 11.1 | 0.5 | 4 | 1 | 11.5 | 0.8 | 7 | 5 | 12.1 | 1.1 | 9 | 10 | 12 | 3 | 29 | 6 | 11 |
| Th | 5.6 | 0.4 | 7 | -4 | 5.3 | 0.2 | 3 | -10 | 5.8 | 0.5 | 8 | -2 | 5 | 1 | 22 | -17 | 5.9 |
| U | 1.6 | 0.1 | 8 | -8 | 1.7 | 0.3 | 15 | -1 | 1.7 | 0.3 | 16 | 3 | 1.4 | 0.3 | 22 | -15 | 1.69 |

Table 3

# Table 4

| Element | 15 micron mean (n=6) | 1 | RSD % | RDRV % | 10 micron mean (n=6) | 1 | RSD % | RDRV % | 8 micron mean (n=6) | 1 | RSD % | RDRV % | 5 micron mean (n=6) | 1 | RSD % | RDRV % | REF |
|---|---|---|---|---|---|---|---|---|---|---|---|---|---|---|---|---|---|
| V | 410 | 16 | 4 | -4 | 401 | 15 | 4 | -6 | 421 | 20 | 5 | -1 | 427 | 33 | 8 | 1 | 425 |
| Mn | 1557 | 72 | 5 | 0 | 1518 | 70 | 5 | -2 | 1585 | 90 | 6 | 2 | 1548 | 146 | 9 | 0 | 1550 |
| Rb | 48 | 3 | 6 | 0 | 48 | 1 | 3 | 1 | 51 | 5 | 9 | 6 | 51 | 7 | 13 | 7 | 48 |
| Sr | 325 | 14 | 4 | -5 | 312 | 16 | 5 | -9 | 321 | 7 | 2 | -6 | 304 | 26 | 9 | -11 | 342 |
| Y | 31 | 2 | 6 | -10 | 30 | 1 | 4 | -10 | 29 | 3 | 9 | -13 | 28 | 2 | 9 | -18 | 34 |
| Zr | 167 | 6 | 4 | -9 | 166 | 9 | 6 | -10 | 160 | 11 | 7 | -13 | 153 | 11 | 7 | -17 | 184 |
| Nb | 11.1 | 0.4 | 3 | -9 | 11 | 1 | 9 | -11 | 10.2 | 0.9 | 8 | -17 | 10.4 | 0.8 | 8 | -16 | 12.3 |
| Ba | 617 | 36 | 6 | -10 | 604 | 20 | 3 | -12 | 620 | 30 | 5 | -9 | 556 | 50 | 9 | -19 | 683 |
| La | 24.2 | 1.6 | 6 | -2 | 23.0 | 1.4 | 6 | -7 | 25 | 2 | 7 | 0 | 22 | 2 | 9 | -11 | 24.7 |
| Ce | 48 | 2 | 5 | -10 | 48.0 | 1.7 | 4 | -10 | 49 | 2 | 3 | -8 | 43 | 2 | 4 | -19 | 53.3 |
| Pr | 6.0 | 0.2 | 3 | -10 | 5.8 | 0.4 | 7 | -13 | 5.6 | 0.5 | 10 | -17 | 5.5 | 0.6 | 11 | -18 | 6.7 |
| Nd | 26.8 | 1.1 | 4 | -7 | 26 | 2 | 8 | -9 | 27 | 3 | 10 | -8 | 21 | 6 | 28 | -26 | 28.9 |
| Sm | 5.9 | 0.5 | 9 | -10 | 5.9 | 0.7 | 11 | -10 | 6.0 | 0.8 | 13 | -9 | 6 | 2 | 34 | -7 | 6.59 |
| Eu | 2.1 | 0.2 | 10 | 6 | 1.9 | 0.3 | 14 | -2 | 2.1 | 0.7 | 32 | 4 | 2.2 | 0.3 | 15 | 13 | 1.97 |
| Gd | 6.5 | 0.4 | 7 | -3 | 6.2 | 0.6 | 9 | -8 | 6 | 1 | 21 | -17 | 7 | 1 | 20 | 3 | 6.71 |
| Tb | 0.96 | 0.08 | 8 | -6 | 0.9 | 0.2 | 19 | -10 | 0.9 | 0.2 | 17 | -9 | 0.8 | 0.1 | 19 | -26 | 1.02 |
| Dy | 5.8 | 0.4 | 6 | -10 | 5.7 | 0.5 | 9 | -11 | 6 | 1 | 15 | -1 | 5 | 1 | 27 | -28 | 6.44 |
| Ho | 1.1 | 0.1 | 10 | -10 | 1.1 | 0.2 | 17 | -10 | 1.2 | 0.2 | 14 | -6 | 1.1 | 0.3 | 25 | -15 | 1.27 |
| Er | 3.4 | 0.2 | 6 | -8 | 3.3 | 0.1 | 3 | -10 | 3.0 | 0.5 | 17 | -19 | 2.7 | 0.2 | 7 | -27 | 3.7 |
| Tm | 0.48 | 0.05 | 10 | -7 | 0.51 | 0.07 | 14 | 0 | 0.6 | 0.1 | 26 | 10 | 0.5 | 0.3 | 57 | -10 | 0.51 |
| Yb | 3.1 | 0.2 | 5 | -9 | 3.2 | 0.7 | 22 | -6 | 2.9 | 0.5 | 19 | -13 | 2.8 | 0.7 | 24 | -16 | 3.39 |
| Lu | 0.49 | 0.05 | 10 | -4 | 0.48 | 0.09 | 19 | -5 | 0.4 | 0.2 | 38 | -11 | 0.5 | 0.1 | 28 | -11 | 0.503 |
| Hf | 4.6 | 0.5 | 10 | -4 | 4.4 | 0.4 | 10 | -10 | 3.9 | 0.7 | 17 | -19 | 3 | 1 | 21 | -28 | 4.84 |
| Ta | 0.70 | 0.05 | 7 | -4 | 0.69 | 0.05 | 7 | -12 | 0.6 | 0.1 | 22 | -19 | 0.6 | 0.2 | 37 | -25 | 0.78 |
| Pb | 10.8 | 0.8 | 7 | -1 | 10 | 1 | 10 | -5 | 12 | 1 | 11 | 5 | 11 | 3 | 30 | -2 | 11 |
| Th | 5.5 | 0.5 | 9 | -7 | 5.2 | 0.4 | 8 | -12 | 5.3 | 0.5 | 10 | -10 | 5 | 1 | 22 | -22 | 5.9 |
| U | 1.5 | 0.1 | 7 | -10 | 1.5 | 0.1 | 9 | -11 | 1.6 | 0.2 | 11 | -3 | 1.4 | 0.3 | 26 | -19 | 1.69 |

Table 4

Table 5

| Sample | SiO$_2$ | TiO$_2$ | Al$_2$O$_3$ | MgO | FeO | MnO | CaO | Na$_2$O | K$_2$O | P$_2$O$_5$ | V | Mn | Rb | Sr | Y | Zr | Nb | Ba |
|---|---|---|---|---|---|---|---|---|---|---|---|---|---|---|---|---|---|---|
| E37-a | 66.90 | 0.80 | 14.34 | 0.56 | 5.73 | 0.18 | 1.72 | 5.72 | 3.90 | 0.16 | 54 | 1720 | 49 | 123 | 66 | 502 | 66 | 367 |
| E37-b | 67.12 | 0.88 | 14.31 | 0.54 | 5.56 | 0.19 | 1.83 | 5.63 | 3.75 | 0.19 | 38 | 1400 | 49 | 125 | 66 | 497 | 64 | 359 |
| E37-c | 60.47 | 1.72 | 14.42 | 1.57 | 8.85 | 0.25 | 3.97 | 5.83 | 2.40 | 0.53 | 24 | 1412 | 44 | 206 | 45 | 373 | 55 | 353 |
| E37-d | 67.44 | 0.90 | 14.13 | 0.55 | 5.39 | 0.20 | 1.68 | 5.73 | 3.75 | 0.24 | 11 | 980 | 72 | 96 | 62 | 668 | 93 | 476 |
| E37-e | 66.56 | 0.83 | 14.36 | 0.69 | 5.67 | 0.17 | 1.92 | 5.91 | 3.70 | 0.18 | 12 | 1066 | 67 | 73 | 64 | 682 | 93 | 447 |
| E37-f | 66.24 | 0.96 | 14.61 | 0.70 | 5.63 | 0.22 | 1.68 | 5.88 | 3.85 | 0.23 | 13 | 1060 | 67 | 178 | 62 | 650 | 84 | 418 |
| E37-g | 66.48 | 0.86 | 14.49 | 0.70 | 5.67 | 0.10 | 1.69 | 5.78 | 3.94 | 0.31 | 12 | 867 | 72 | 61 | 56 | 660 | 91 | 398 |
| E37-h | 61.20 | 1.45 | 13.79 | 1.47 | 9.59 | 0.20 | 3.31 | 5.41 | 3.15 | 0.43 | 22 | 1410 | 46 | 187 | 51 | 427 | 64 | 405 |
| E37-i | 66.92 | 0.87 | 14.40 | 0.64 | 5.43 | 0.22 | 1.89 | 5.86 | 3.60 | 0.16 | 11 | 929 | 68 | 87 | 62 | 652 | 89 | 420 |
| E37-l | 66.80 | 0.94 | 14.67 | 0.56 | 5.48 | 0.23 | 1.74 | 5.75 | 3.54 | 0.28 | 12 | 1072 | 69 | 81 | 62 | 672 | 98 | 456 |
| E37-m | 69.54 | 0.86 | 13.94 | 0.36 | 4.19 | 0.02 | 1.26 | 5.30 | 4.37 | 0.15 | 13 | 1060 | 67 | 123 | 64 | 623 | 88 | 500 |
| E37-n | 66.40 | 0.82 | 14.48 | 0.65 | 5.35 | 0.24 | 1.97 | 5.90 | 3.83 | 0.35 | 12 | 974 | 66 | 138 | 62 | 634 | 88 | 469 |
| E37-o | 67.30 | 0.82 | 14.27 | 0.60 | 5.37 | 0.23 | 1.51 | 5.81 | 3.88 | 0.21 | 23 | 1230 | 68 | 89 | 63 | 647 | 94 | 434 |
| E37-p | 66.54 | 0.88 | 14.34 | 0.63 | 5.86 | 0.15 | 1.91 | 6.00 | 3.54 | 0.15 | 37 | 1470 | 57 | 157 | 58 | 638 | 84 | 458 |

| Sample | La | Ce | Pr | Nd | Sm | Eu | Gd | Tb | Dy | Ho | Er | Tm | Yb | Lu | Hf | Ta | Pb | Th | U |
|---|---|---|---|---|---|---|---|---|---|---|---|---|---|---|---|---|---|---|---|
| E37-a | 48 | 110 | 13 | 56 | 13 | 3.3 | 14 | 2.2 | 13 | 2.9 | 8.0 | 1.1 | 6.8 | 0.97 | 12 | 3.8 | 4.7 | 6.0 | 2.2 |
| E37-b | 55 | 138 | 16 | 70 | 16 | 3.3 | 14 | 2.2 | 13 | 2.5 | 6.3 | 0.9 | 7.7 | 1.25 | 12 | 3.7 | 3.2 | 5.6 | 1.9 |
| E37-c | 43 | 94 | 11 | 46 | 10 | 3.1 | 10 | 1.4 | 9 | 1.9 | 4.3 | 0.6 | 4.3 | 0.54 | 9 | 3.1 | 4.5 | 4.8 | 1.8 |
| E37-d | 60 | 136 | 16 | 57 | 14 | 2.4 | 10 | 1.8 | 13 | 2.1 | 7.7 | 1.1 | 6.7 | 0.78 | 15 | 4.9 | 7.5 | 8.2 | 2.9 |
| E37-e | 64 | 139 | 17 | 66 | 15 | 2.7 | 14 | 1.9 | 15 | 2.7 | 7.2 | 1.0 | 7.2 | 0.99 | 17 | 5.5 | 8.0 | 9.9 | 2.8 |
| E37-f | 59 | 129 | 14 | 57 | 13 | 3.1 | 11 | 1.6 | 11 | 2.2 | 6.1 | 0.9 | 5.0 | 0.76 | 13 | 4.7 | 7.5 | 7.2 | 2.6 |
| E37-g | 60 | 129 | 15 | 57 | 11 | 2.7 | 10 | 1.7 | 11 | 2.1 | 5.5 | 1.0 | 6.9 | 0.89 | 15 | 5.3 | 6.3 | 8.4 | 3.4 |
| E37-h | 47 | 105 | 13 | 51 | 13 | 3.9 | 10 | 1.5 | 9 | 2.0 | 4.9 | 0.7 | 4.1 | 0.56 | 10 | 3.7 | 4.4 | 5.5 | 2.0 |
| E37-i | 61 | 133 | 15 | 64 | 13 | 2.3 | 12 | 1.8 | 13 | 2.1 | 6.5 | 1.0 | 6.2 | 1.00 | 15 | 5.5 | 5.8 | 7.8 | 3.0 |
| E37-l | 67 | 150 | 17 | 68 | 12 | 2.6 | 13 | 2.0 | 12 | 2.2 | 6.9 | 0.9 | 5.8 | 0.86 | 15 | 5.5 | 6.7 | 8.6 | 3.5 |
| E37-m | 60 | 129 | 15 | 60 | 14 | 3.0 | 13 | 2.0 | 12 | 2.3 | 6.6 | 1.0 | 5.9 | 0.76 | 14 | 5.3 | 6.4 | 7.3 | 2.8 |
| E37-n | 60 | 131 | 16 | 64 | 13 | 3.0 | 11 | 1.7 | 12 | 2.3 | 7.3 | 1.0 | 6.8 | 0.96 | 15 | 5.2 | 7.2 | 7.7 | 2.9 |
| E37-o | 63 | 138 | 16 | 64 | 14 | 2.4 | 13 | 1.9 | 14 | 2.3 | 6.9 | 1.1 | 7.2 | 0.90 | 14 | 5.0 | 6.0 | 8.1 | 2.9 |
| E37-p | 54 | 124 | 14 | 58 | 10 | 2.9 | 11 | 1.6 | 11 | 2.2 | 5.2 | 0.8 | 5.0 | 0.87 | 14 | 4.6 | 6.8 | 7.4 | 3.4 |

Table 5